\documentclass[12pt]{article}
\usepackage[top = 2 cm, bottom = 3 cm, left = 2.5 cm, right = 1.8 cm]{geometry}
\usepackage{authblk, cite, color, amssymb, amsmath}
\usepackage[colorlinks = true, linkcolor = blue, citecolor = red]{hyperref}
\usepackage[dvipsnames]{xcolor}

\begin{document}

\title{Cosmological Constant from the Entropy Balance Condition}

\author{\bf M. Gogberashvili}
\affil{\small Javakhishvili Tbilisi State University, 3 Chavchavadze Avenue, Tbilisi 0179, Georgia \authorcr
Andronikashvili Institute of Physics, 6 Tamarashvili Street, Tbilisi 0177, Georgia}

\maketitle

\begin{abstract}
In the action formalism variations of metric tensors usually are limited by the Hubble horizon. Contrary, variations of quantum fields should be extended up to the event horizon, which is the real boundary of the spacetime. As the result the entanglement energy of quantum particles across the apparent horizon is missed in the cosmological equations written for the Hubble volume. We identify this missing boundary term with the dark energy density and express it (using the null energy assumption for the finite universe) as the critical density multiplied by the ratio of the Hubble and event horizons radii.

\vskip 1mm

\end{abstract}
\vskip 5mm


Many authors consider so called emergent theories in which gravity is not a fundamental field, but like thermodynamics or hydrodynamics is defined for the matter in bulk \cite{eff-grav-1, eff-grav-2, Verl, Jac, Pad}. One such approach is the thermodynamic model of gravity \cite{Verl, Jac, Pad, Gog-Chut, Gog-Kanat-2, Gog-4, Gog-2, Gog-Kanat-1, Gog-3, Gog-1, Gog-5}, where spacetime emerges from the properties of the 'universal' ensemble of quantum particles. In this approach the entropy, rather than the energy density, plays the crucial role.

Since in General Relativity horizons are unavoidable and horizons block information, entropy and temperature can be introduced for spacetime. One such boundary is the apparent horizon with the radius (for the spatially flat universe),
\begin{equation} \label{R_H}
R_H = \frac 1H \approx 14.5~Gly~,
\end{equation}
where $H \equiv \dot {a}/a$ is the Hubble parameter. For the Hubble volume,
\begin{equation} \label{V_H}
V_H = \frac 43 \pi R_H^3~,
\end{equation}
having the surface area
\begin{equation} \label{A=}
A_H = 4\pi R_H^2~,
\end{equation}
one can associated the temperature \cite{Haw},
\begin{equation} \label{T}
T = \frac {1}{2\pi R_H}~,
\end{equation}
and an entropy \cite{Bek},
\begin{equation} \label{S=A/4}
S_H = \frac {A_H}{4G}~,
\end{equation}
where $G$ is the Newton constant.

The concept of entropy is a powerful tool in thermodynamics, information theory and quantum physics, and allows us to study different aspects of physical systems using a similar mathematical framework. In quantum mechanics a measurement is considered as the interaction of three systems: the quantum object, memory (measurement device) and observer. Then the total entropy of the ensemble of all quantum particles, which is formed by the information, statistical (thermodynamic) and quantum (entanglement) components \cite{Hwan, Song}, can be assumed to be zero \cite{Gog-5}. In this case the universe can always remain in pure state and only allow a unitary time-evolution, as it is suggested by von Neumann's model. Also the 'universal' entropy remains zero at all stages of universe's evolution, while any it subsystem (for example the Hubble volume) has non-zero entropy.

In our previous papers it was demonstrated that 'world ensemble' approach is compatible with the existing field-theoretical descriptions, as the relativistic \cite{Gog-2, Gog-Kanat-1} and quantum \cite{Gog-3} properties are emerging from its properties. Moreover, the model allows us to explain the hierarchy problem in particle physics by the fact that our underlying assumption that any gravitational interaction of two particles involves interactions with all particles of the world ensemble effectively weakens the observed strength of gravity by the factor proportional to the number of particles in the ensemble \cite{Gog-1}.

As it was noted in \cite{Gog-5}, the convenient physical parameter to measure information can be an action, which in most cases is an additive quantity like entropy, also contains positive and negative components and exhibits the unique discrete value, the action quantum $\hbar$ \cite{Ann}. Then the maximum entropy principle in thermodynamics \cite{Jay} and the least action principle in field theory, lead the same formalism. The relation of the classical action of a physical system to the thermodynamic entropy translates the condition of entropy neutrality into the null action principle - the sum of all components of the action for a physical system (including the boundary terms) is zero. One consequence of this principle for the whole universe is the zero-energy condition,
\begin{equation} \label{rho-U=0}
\rho_U = 0~,
\end{equation}
i.e. the total density of all the forms of energy in the universe, $\rho_U$, at any moment of time should be zero. This mean that the universe can emerge without violation of the energy conservation, which appears to be preferable point of view in cosmology \cite{Fey,Haw-2}.


In this paper we want to connect the entanglement energy of quantum particles across the apparent horizon, which is missed in the classical cosmological equations written for the Hubble volume, with the Dark Energy (DE), origin of which remains a mystery \cite{Roos}. It is known that DE can be modelled by Einstein's cosmological constant, for which a typical fit to current observational data gives $\Lambda \approx 3\times 10^{-122}~sec^{-2}$. However, in quantum field theory, the natural value of $\Lambda$ is of the order of unity. This discrepancy is one of the biggest challenges in modern cosmology and fundamental physics \cite{Wein, HEL}.

Alternatives of the introduction of a cosmological constant are that DE arises from the evolution of dynamical fields of an unknown origin, or modifications of General Relativity. In order to distinguish between these hypotheses, a worldwide effort is ongoing to measure the effective equation of state and clustering properties of DE, using wide field cosmological surveys \cite{Nojiri}. For reviews on the DE and theories, see \cite{DE-rev-1, DE-rev-2, DE-rev-3, DE-rev-4} and references therein.

One promising approach for solving the DE puzzle is the Holographic DE model \cite{CKN, Thomas, Hsu, Li}, which is based on the quantum zero-point energy predicted by an effective Quantum Field Theory. The primary model of this kind, which as the IR cut-off uses the apparent horizon of the universe, $R_H$, has serious drawbacks \cite{GHN, Horvat, Gao}, since for the DE density predicts the value:
\begin{equation} \label{HolDE}
\rho_{DE} = 2\pi\rho_c ~,
\end{equation}
where
\begin{equation} \label{rho_c}
\rho_c = \frac {3 H^2}{8\pi G}
\end{equation}
denotes the critical density. As we see the vacuum energy density in Holographic DE model (\ref{HolDE}) is larger than observed, in fact it even exceeds the critical density $\rho_c$. To solve this failure physicists are considering \cite{Wang}: i) interactions between the cosmos sectors; ii) various models for entropies; iii) different from $R_H$ cut-offs.

We note that, based on spacetime thermodynamics, a proper causal boundary of the classical spacetime is its apparent horizon \cite{FRW-bound-1, FRW-bound-2}, meaning that the metric fluctuations are bounded by $R_H$ and also that thermodynamics laws are satisfied on this boundary \cite{Termo-bound-1, Termo-bound-2}. Moreover, the event horizon in the context of cosmology as well as in the context of a black hole is always defined globally, as the causal structure of spacetime is a global thing (see more discussions in \cite{Li}). From the other hand, the quantum fluctuations of matter fields should be limited not by $R_H$, but by the event horizon,
\begin{equation} \label{R_e}
R_e = \int_1^\infty \frac {da}{a^2 H(a)} \approx 16.7~Gly~,
\end{equation}
which represents a real boundary of spacetime. Then the entanglement energy of quantum particles across the apparent horizon $R_H$, which is defined as disturbed vacuum energy due to the presence of a boundary \cite{MSK}, is missed in the cosmological equations written for the Hubble volume and can be taken into account by introduction of a boundary term. It was found that the perfect fluid of entanglement has a negative pressure \cite{LLK} and can be interpreted as the origin of DE. The terms corresponding to entanglement across the horizon should disappear in the equation of state of classical fields. Indeed, for the light-like geodesics, which describe the Hubble horizon in the spatially flat universe, the Einstein equations written in the form of the first law does not contains the DE terms.

In thermodynamic approach the Einstein equations can be derived by combining general thermodynamic considerations with the equivalence principle and be written as a single scalar relation \cite{Jac, Pad, Gog-Chut},
\begin{equation} \label{Termo-Ein}
\left( R_{\mu \nu} - \frac 12 g_{\mu\nu}R \right) u^\mu u^\nu = 8\pi G T_{\mu \nu} u^\mu u^\nu~.
\end{equation}
This equation involve additional vector field $u^\nu$, but contains all information content of the ordinary tensorial Einstein equations, because it is demanded that it holds for all $u^\nu$. In addition, if one assumes that $u^\nu$ is an orthogonal to the observers horizon null vector field \cite{Pad, Gog-Chut},
\begin{equation} \label{l^2=0}
g_{\mu\nu}u^\mu u^\nu = 0~,
\end{equation}
in the obtained from (\ref{Termo-Ein}) tensorial Einstein equations,
\begin{equation} \label{Ein}
R_{\mu \nu} - \frac 12 g_{\mu\nu}R = 8\pi G \left(T_{\mu \nu} + g_{\mu\nu}\Lambda\right)~,
\end{equation}
$\Lambda$ arises as an integration constant, which is not connected with the large constant vacuum energy terms in matter Lagrangians and needs to be fixed using an extra physical principle. For example, since we know that information is a physical entity \cite{Land, Bril}, it can be identified with the amount of information accessible to an eternal observer at the horizon \cite{Gog-Chut}, or with the energy of collective gravitational interactions of all particles in the finite universe \cite{Gog-Kanat-2, Gog-4}.

The equation (\ref{Termo-Ein}) has the natural interpretation as the balance of gravitational and matter heat densities, where the right hand side represents the matter heat density, in the spirit of the first law of thermodynamics. This is obvious, for example, for the case of ideal fluid using the classical Gibbs-Duhem relation,
\begin{equation}\label{Gibbs-Duhem}
T_{\mu \nu} u^\mu u^\nu \to \rho + p = \frac {TS_m}{V} ~,
\end{equation}
where $T$ is the temperature and $S_m$ is the entropy of the matter in the volume $V$.

Connections of the cosmological constant with the boundary conditions and generalized equations of state can be demonstrated also from the cosmological equations. For a homogeneous, isotropic and flat universe ($k = 0$) there are two independent Friedmann equations with the cosmological term $\Lambda$:
\begin{eqnarray} \label{FRW}
H^2 &=& \frac {8\pi G}{3} \rho + \frac 13 \Lambda~ \nonumber \\
\dot{H} + H^2 \equiv \frac {\ddot a}{a} &=& - \frac {4\pi G}{3} (\rho + 3 p) + \frac 13 \Lambda~.
\end{eqnarray}
It is known that using the first equation of (\ref{FRW}) the second equation can be re-expressed without $\Lambda$,
\begin{equation} \label{dotH}
\dot{H} = - 4\pi G (\rho + p)~.
\end{equation}
From the system (\ref{FRW}) one can obtain also the matter energy-momentum conservation condition,
\begin{equation} \label{dT}
\partial_\nu T^{\nu\mu} = 0 ~,
\end{equation}
which leads to:
\begin{equation} \label{dotRho}
\dot{\rho} = - 3 H (\rho + p)~,
\end{equation}
Due to the presence of derivatives in (\ref{dT}) the cosmological constant $\Lambda$ does not appears in (\ref{dotRho}) as well.

So, if instead of (\ref{FRW}), one will choice (\ref{dotH}) and (\ref{dotRho}) as the independent system of cosmological equations, $\Lambda$ obtains the role of integration constant which can be fixed from an equation of state. Indeed, combining (\ref{dotH}) with (\ref{dotRho}) and integrating over the time we have,
\begin{equation} \label{balance}
H^2 = \frac {8\pi G}{3} \rho + C~,
\end{equation}
where the cosmological term re-appears in the form of an arbitrary constant $C$, which should be chosen as,
\begin{equation} \label{C}
C = \frac 13 \Lambda~,
\end{equation}
in order to obtain the first Friedmann equation in (\ref{FRW}).

In General Relativity, as we know, $\Lambda$ is a true constant and is not seen to evolve. However, an expanding universe is not expected to have a static vacuum energy density and there is the possibility that $\Lambda$ is actually a time dependent quantity. In order to implement the notion of a smoothly evolving vacuum energy density is not necessary to introduce ad hoc scalar fields, as usually done in quintessence formulations. Let us mention a dynamical approach based on extending the variational principle by promoting $\Lambda$ from being a parameter to a field \cite{Shaw-Bar}. In this modified variational approach the resulting history is indistinguishable from General Relativity with a constant value of $\Lambda$  put in by hand at the right value at each observational time.


Let us estimate the entropy input in a region as the sum of the entropy flux (entropy received per unit surface) transferred through the boundary, and the entropy supplied by internal sources (entropy generated per unit volume). If we neglect the entropy supplied by internal sources, then according the Second Law of thermodynamics, the time derivative of the entropy contained within the volume, $S$, is equal to the flux of the matter entropy, $S_m$, through the boundary $A$,
\begin{equation} \label{dS=}
\frac {dS}{dt} = S_m A ~.
\end{equation}
Using the relations of the type (\ref{S=A/4}) and (\ref{Gibbs-Duhem}) the equation (\ref{dS=}) takes the form:
\begin{equation} \label{}
\frac {1}{4G}\frac {dA}{dt} = \frac{\rho + p}{T} A~.
\end{equation}
For the Hubble volume, taking into account (\ref{A=}) and (\ref{T}), this equation leads to one of the cosmological equations (\ref{dotH}). Then combined with the energy conservation equation (\ref{dotRho}), and integrating over time we find again (\ref{balance}), where $C$ corresponds to some hidden amount of energy.

For the finite universe limited by the event horizon, $R_e$, the energy balance condition (\ref{balance}) should obtain the form:
\begin{equation} \label{}
\frac {1}{R_e^2} = \frac {8\pi G}{3}\rho_U~,
\end{equation}
without the boundary term. Then the zero energy condition (\ref{rho-U=0}) allows us to fix $C$ in (\ref{balance}),
\begin{equation} \label{}
C = \frac 13 \Lambda = \frac {1}{R_e^2}~.
\end{equation}
Therefore, using (\ref{R_H}) and (\ref{R_e}) the DE density can be estimated as,
\begin{equation} \label{DE}
\rho_{DE} = \rho_c \frac {C}{H^2} = \rho_c \frac {R_H^2}{R_e^2} = 0.75 \rho_c ~.
\end{equation}
which is very close to the observed value of the DE.


To conclude, in this paper we estimate the DE density within thermodynamic model of gravity using the null energy condition - the total energy of the universe inside its event horizon is zero. We notice that in the action formalism variations of metric tensor should be limited by the Hubble horizon, which represents a causal boundary of classical spacetime. Contrary, variations of quantum fields are limited by the event horizon, which is the real boundary of the spacetime. Then the entanglement energy of quantum particles across the apparent horizon is missed in the cosmological equations written for the Hubble volume. We identify this entanglement density with the DE, which can be introduced as a boundary term in the cosmological equations. In our model the DE density appears to be equal to the critical density reduced by the ratio of the squares of the Hubble and event horizons radii (\ref{DE}), having the value in good agreement with the observational data.


\end{document}